\documentclass{optica-article}

\journal{opticajournal} % for Optics Express / Optical Materials Express

\articletype{Research Article}

\usepackage{lineno}
\usepackage{amsmath}
\usepackage{bm}

\usepackage{graphicx}
\usepackage{subcaption}
\usepackage{float}

\usepackage{booktabs}
\usepackage{array}
\usepackage{multirow}

\usepackage{siunitx}
\newcommand{\ReS}{ReS\textsubscript{2}}
\newcommand{\MoO}{$\alpha$-MoO\textsubscript{3}}
\newcommand{\lres}{\lambda_{\mathrm{res}}}
\newcommand{\exx}{\varepsilon_{xx}}
\newcommand{\eyy}{\varepsilon_{yy}}
\newcommand{\ezz}{\varepsilon_{zz}}
\newcommand{\gr}{\gamma_{r}}
\newcommand{\ga}{\gamma_{a}}
\newcommand{\Pabs}{P_{\mathrm{abs}}}

\begin{document}
%% ─────────────────────────────────────────────────────────────────────────────

\title{Polarization-Selective Near-Perfect Absorption via Mie-Type Resonance
in van der Waals Anisotropic \ReS/\MoO/Au Heterostructure}

\author{Shoumik Debnath,\authormark{1} and Sudipta Saha\authormark{1,*}}

\address{\authormark{1}Department of Electrical and Electronic Engineering,
Bangladesh University of Engineering and Technology (BUET),
Dhaka 1205, Bangladesh}

\email{\authormark{*}sudiptasaha@ari.buet.ac.bd}

\begin{abstract*}
We investigate polarization-selective absorption in a visible-wavelength heterostructure consisting of a ReS$_2$ stripe grating, an $\alpha$-MoO$_3$ spacer, and an Au back-reflector using finite-difference time-domain simulations. For an optimized geometry with a grating period of \SI{500}{\nano\meter}, stripe width of \SI{250}{\nano\meter}, and ReS$_2$ thickness of \SI{80}{\nano\meter}, the structure exhibits near-unity absorption of 99.99\% at \SI{650.5}{\nano\meter} under TE-polarized illumination. The resonant field is concentrated near the outer edges of the ReS$_2$ stripe, while absorption power density is localized in the same region, consistent with a localized edge mode. The absorption response depends strongly on polarization, producing a TE--TM resonance separation of \SI{16.2}{\nano\meter}. Replacing either the biaxial ReS$_2$ layer or the anisotropic $\alpha$-MoO$_3$ spacer with isotropic equivalents substantially modifies the spectral response and reduces the polarization-dependent wavelength separation. In addition, rotating the crystal orientation of the ReS$_2$/$\alpha$-MoO$_3$ stack shifts both the resonance wavelength and peak absorption without changing the device geometry. The results show that the combination of anisotropic resonator and spacer layers provides an effective means of controlling resonant absorption and polarization selectivity in van der Waals photonic structures.
\end{abstract*}

%% ─────────────────────────────────────────────────────────────────────────────
\section{Introduction}
%% ─────────────────────────────────────────────────────────────────────────────

Polarization-selective absorbers are important for applications including
Stokes polarimetry, hyperspectral imaging, and coherent optical sensing~\cite{Kuznetsov2016,Limonov2017}.
Achieving strong absorption for one polarization while suppressing the
orthogonal polarization enables compact passive devices for polarization
discrimination.
Near-perfect narrowband absorption can be obtained through critical
coupling~\cite{Landy2008,Fan2003}, where the radiative decay rate of a
resonant mode matches its intrinsic absorption rate.
This condition has been demonstrated in metal--dielectric--metal
(Salisbury-screen) structures across a wide spectral range extending
from microwave to near-infrared wavelengths~\cite{Landy2008,Kats2013}.
However, visible-wavelength absorbers that rely on intrinsic material
anisotropy rather than geometric symmetry breaking remain relatively
unexplored.

Transition metal dichalcogenides (TMDs) have been widely investigated for
visible nanophotonics~\cite{Wilson1969,Mak2010,Wang2018}.
Multilayer flakes exhibit in-plane refractive indices of
$n \gtrsim 3.5$ across much of the visible spectrum, driven by strong excitonic
oscillator strengths near the A-, B-, and C-exciton
resonances~\cite{Munkhbat2022,Laturia2018}.
These high indices enable strongly subwavelength electromagnetic confinement that is
unavailable in conventional grating materials such as TiO\textsubscript{2}
($n \approx 2.3$) or even silicon above its bandgap~\cite{Kuznetsov2016}.
Patterned TMD nanostructures support localized Mie dipole resonances of both electric
and magnetic character~\cite{Verre2019,Green2020}, and the van der Waals
layered structure imparts a large out-of-plane birefringence
($\Delta n \gtrsim 1.3$) across the TMD family, confirmed by near-field microcrystal imaging~\cite{Hu2017,Ermolaev2021,Munkhbat2022}.
All-TMD nanophotonic integration exploiting these properties has been
demonstrated in several system configurations~\cite{Ling2021}.

Among semiconducting TMDs, rhenium disulfide (ReS\textsubscript{2}) occupies a
singular position.
Unlike the $2H$-phase family (MoS\textsubscript{2}, WS\textsubscript{2}, and their
selenide analogues), whose hexagonal symmetry enforces in-plane isotropy
$\exx = \eyy$, ReS\textsubscript{2} adopts the $1T''$ distorted-octahedral
structure~\cite{Aslan2016,Wilson1969}.
Re--Re covalent chains along the $b$-axis break the six-fold rotational symmetry
and reduce the effective crystal symmetry to triclinic, producing a full in-plane
biaxial dielectric tensor: $\exx(\lambda) \neq \eyy(\lambda) \neq \ezz(\lambda)$
throughout the visible and near-infrared~\cite{Aslan2016,Munkhbat2022,Shubnic2020}.
Related rhenium dichalcogenides share this $P\bar{1}$ symmetry and are identified
as among the highest-index biaxially anisotropic semiconductors
available~\cite{Shubnic2020,Mikhin2023}.
Experimental manifestations of ReS\textsubscript{2} anisotropy include
linearly polarized in-plane excitons~\cite{Aslan2016},
self-hybridized polaritons~\cite{Gogna2020},
and polarization-sensitive photodetection using ReS\textsubscript{2}-based Mie
structures~\cite{Yan2023}.
Near-field waveguide-mode spectroscopy on thin ReS\textsubscript{2} flakes
(30--59~nm) has confirmed that the biaxial anisotropy is spatially
uniform at the nanoscale and bulk-consistent well below
\SI{100}{\nano\meter}~\cite{Mooshammer2022}.

$\alpha$-Molybdenum trioxide ($\alpha$-MoO\textsubscript{3}) is an orthorhombic van
der Waals crystal (Pbnm space group) with three crystallographically and optically
inequivalent principal axes $\alpha$~([100]), $\beta$~([010]),
and $\gamma$~([001])~\cite{Lajaunie2013,Ma2018}.
The in-plane refractive indices form the hierarchy $n_{\alpha} < n_{\beta} < n_{\gamma}$
throughout the visible, with values of approximately 2.13, 2.48, and 2.72 at
\SI{650}{\nano\meter}.
Anisotropy in \MoO{} originates from a strongly inhomogeneous spatial distribution
of empty states localized around terminal oxygen sites, an electronic effect that
persists from bulk to the few-layer limit~\cite{Lajaunie2013}.
While \MoO{} has attracted intense interest for mid-infrared phonon-polariton
propagation~\cite{Ma2018}, its use as an anisotropic spacer layer in visible absorbers
has been explored in nanoribbon metamaterial geometries~\cite{Tang2021}, but without
coupling to a biaxial grating material whose resonance carries intrinsic polarization
character.

Prior TMD absorber studies have employed either isotropic film stacks or
nanodisk arrays where polarization contrast arises from geometric meta-atom
asymmetry~\cite{Verre2019,Green2020,Munkhbat2019}.
The combination of a biaxial \ReS{} grating with an anisotropic \MoO{} spacer,
designed to exploit \emph{dual material anisotropy} for critical-coupled
polarization-selective absorption, has not been previously reported.

In this study, we present an FDTD study of the heterostructure
Air$\,|\,$\ReS{} grating$\,|\,$\MoO{}$\,|\,$Au
in Lumerical.
A \ReS{} stripe grating (period \SI{500}{\nano\meter}, width \SI{250}{\nano\meter},
thickness \SI{80}{\nano\meter}) supports a localized Mie-type edge resonance near
\SI{650}{\nano\meter}.
The \MoO/Au back-reflector cavity brings the structure to critical coupling,
giving $A \approx 99.99\%$ under TE illumination.
The TE--TM splitting of \SI{16.2}{\nano\meter} arises from the biaxial permittivity
of ReS\textsubscript{2}; two control simulations isolate the independent
contributions of ReS\textsubscript{2} biaxiality and \MoO{} anisotropy to this
splitting, and a crystal-orientation sweep demonstrates materials-level spectral
tunability without any geometric modification.

%% ─────────────────────────────────────────────────────────────────────────────
\section{Device Structure and Simulation Methodology}
%% ─────────────────────────────────────────────────────────────────────────────

\subsection{Device architecture}

\begin{figure}[t]
  \centering
  \includegraphics[width=\textwidth]{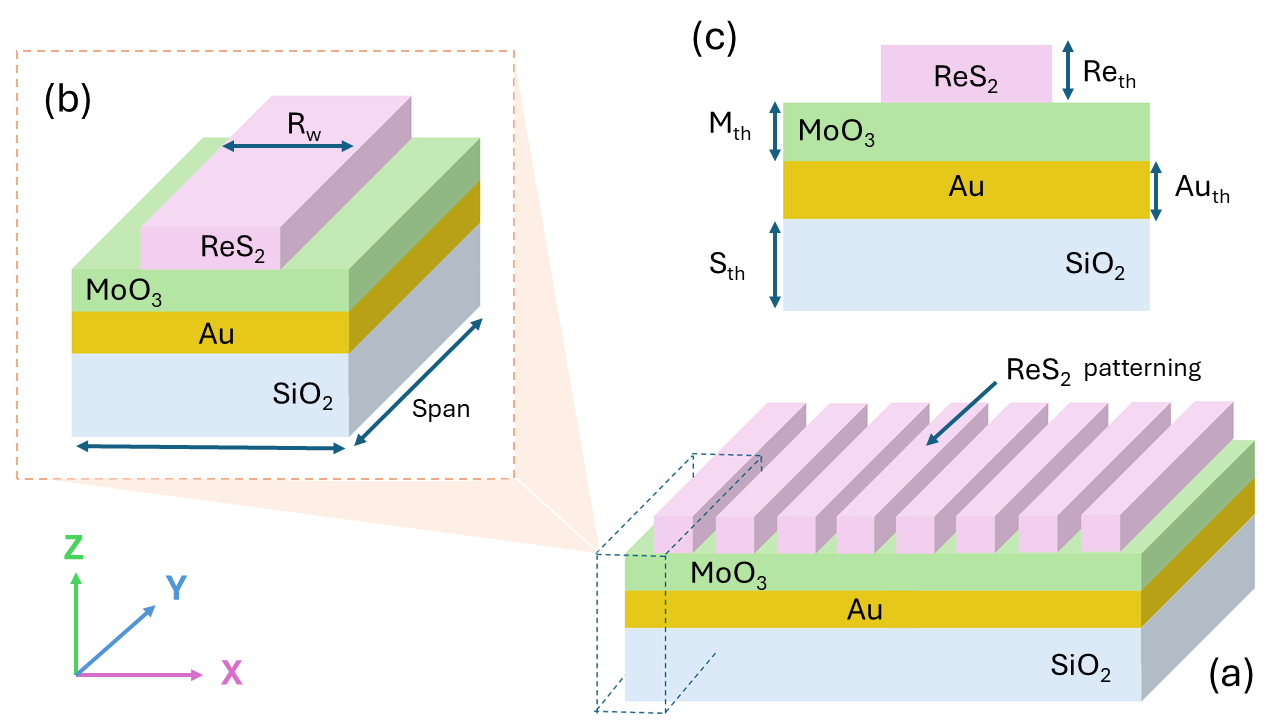}
  \caption{Schematic of the \ReS/\MoO/Au absorber.
    (a)~Full layer stack showing material sequence, (b)~single unit cell with
    labeled stripe width $R_w$ and period (Span), (c)~cross-sectional view with
    layer thicknesses.
    A normally incident plane wave propagates along $-z$; the $x$-axis
    defines the TE ($\bm{E} \parallel \hat{x}$) direction.}
  \label{fig:device}
\end{figure}

The absorber stack is shown schematically in Fig.~\ref{fig:device}.
From top to bottom it consists of air, a periodic \ReS{} stripe grating,
a continuous \MoO{} film, a \SI{150}{\nano\meter} Au back-reflector, and a
SiO\textsubscript{2} substrate.
The \ReS{} stripes have grating period $\Lambda = \SI{500}{\nano\meter}$,
stripe width $w = \SI{250}{\nano\meter}$ (duty cycle 0.5),
and thickness $t_{\mathrm{ReS_2}} = \SI{80}{\nano\meter}$.
The \MoO{} spacer thickness was swept from 100 to \SI{300}{\nano\meter}
and optimized for maximum TE absorption at the resonance wavelength;
all results below correspond to the optimized geometry.
The Au layer is opaque across the visible, so the structure functions as a one-port
reflector ($T = 0$, $A = 1 - R$).
Geometric parameters are compiled in Table~\ref{tab:geometry}.

\begin{table}[htbp]
  \centering
  \caption{Geometric parameters of the absorber.}
  \label{tab:geometry}
  \smallskip
  \begin{tabular}{lc}
    \toprule
    Parameter & Value \\
    \midrule
    Grating period $\Lambda$ & \SI{500}{\nano\meter} \\
    \ReS{} stripe width $w$ & \SI{250}{\nano\meter} \\
    Duty cycle $w/\Lambda$ & 0.5 \\
    \ReS{} thickness $t_{\mathrm{ReS_2}}$ & \SI{80}{\nano\meter} \\
    \MoO{} thickness $t_{\mathrm{MoO_3}}$ & \SI{300}{\nano\meter} \\
    Au thickness $t_{\mathrm{Au}}$ & \SI{150}{\nano\meter} \\
    Substrate & SiO\textsubscript{2} \\
    Incident medium & Air \\
    \bottomrule
  \end{tabular}
\end{table}

\subsection{Material optical constants}

\begin{figure}[htbp]
  \centering
  \includegraphics[width=\textwidth]{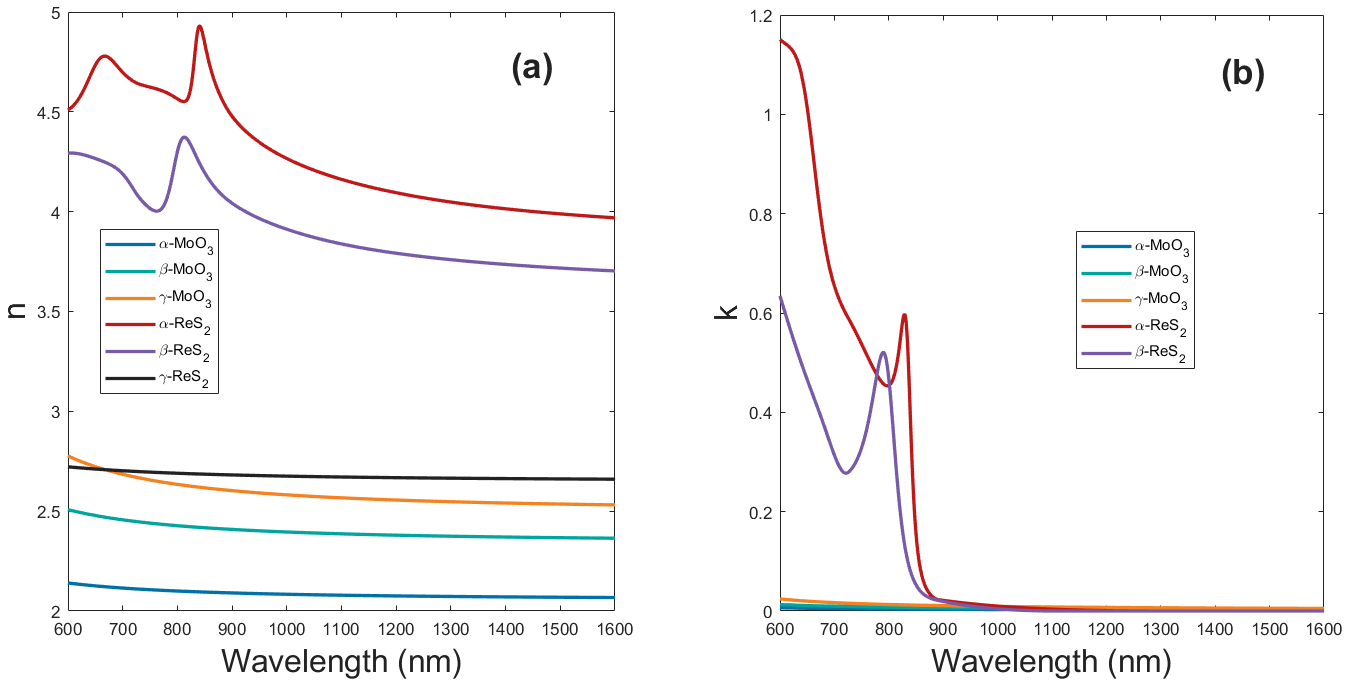}
  \caption{Real ($n$) and imaginary ($k$) parts of the complex refractive index.
    Left panels: \ReS{} biaxial tensor components $xx$, $yy$, $zz$ from
    Ref.~\cite{Munkhbat2022}.
    Right panels: \MoO{} principal axes $\alpha$, $\beta$, $\gamma$ from
    Ref.~\cite{Lajaunie2013}.}
  \label{fig:nk}
\end{figure}

\paragraph{ReS\textsubscript{2}.}
We use the biaxial permittivity tensor
$\bm{\varepsilon} = \mathrm{diag}(\exx,\,\eyy,\,\ezz)$
extracted by Munkhbat et al.\ from multisample Mueller-matrix
spectroscopic ellipsometry on mechanically exfoliated
ReS\textsubscript{2} flakes spanning 200--\SI{600}{\nano\meter}
thickness~\cite{Munkhbat2022}.
The tensor components are parameterized with Tauc--Lorentz oscillator models
and satisfy $\mathrm{Re}(\exx) \neq \mathrm{Re}(\eyy) \neq \mathrm{Re}(\ezz)$
across the visible range (Fig.~\ref{fig:nk}).
At $\lambda = \SI{650}{\nano\meter}$, the principal in-plane refractive indices are
approximately $n_{xx} \approx 4.0$ and $n_{yy} \approx 3.7$.

Three considerations establish the suitability of the Munkhbat tensor for the
\SI{80}{\nano\meter} device layer.
First, \SI{80}{\nano\meter} corresponds to approximately 120 monolayers
(monolayer thickness $\approx \SI{0.68}{\nano\meter}$)~\cite{Aslan2016,Wilson1969},
placing the device layer firmly in the bulk optical regime by any reasonable
criterion.
Second, the $1T''$ crystal structure of ReS\textsubscript{2} exhibits
negligibly weak van der Waals interlayer coupling, in sharp contrast to the $2H$
TMDs (MoS\textsubscript{2}, WS\textsubscript{2}) where strong interlayer
hybridization causes large thickness-dependent shifts in optical constants from
monolayer to bulk~\cite{Mak2010,Laturia2018}.
In ReS\textsubscript{2}, each layer interacts essentially as an isolated slab,
so the biaxial tensor is consistent from the few-layer limit
upward~\cite{Aslan2016,Mooshammer2022}.
Third, the multisample ellipsometric methodology of Ref.~\cite{Munkhbat2022}
simultaneously fits flakes across a 3$\times$ thickness range specifically to
deconvolve thickness-dependent artifacts and extract intrinsic, thickness-independent
bulk tensor components.
A remaining reviewer concern is that the ellipsometric validation range
(200--600~nm) does not directly include \SI{80}{\nano\meter}, is addressed by
the second and third points: the extraction methodology yields intrinsic constants
by design, and near-field waveguide-mode measurements on ReS\textsubscript{2}
flakes in the 30--\SI{59}{\nano\meter} range independently confirm that the
biaxial anisotropic response is bulk-consistent below \SI{100}{\nano\meter}~\cite{Mooshammer2022}.
One sentence suffices as the pre-emptive statement in the manuscript:
\emph{``80~nm corresponds to $\approx$120 monolayers, placing the \ReS{} device
layer firmly in the bulk optical regime where the extracted dielectric tensor
is thickness-independent.''}\vspace{2pt}

\paragraph{$\alpha$-MoO\textsubscript{3}.}
We use the anisotropic dielectric function of Lajaunie et al.~\cite{Lajaunie2013},
derived by combining valence electron energy-loss spectroscopy with
random-phase approximation ab initio calculations including local-field effects
along all three principal axes.
The $\alpha$, $\beta$, $\gamma$ components are mapped to the simulation
coordinate system as described in Section~\ref{sec:fdtd}.
Au optical constants follow Palik~\cite{Palik98}.
SiO\textsubscript{2} is treated as non-dispersive ($n = 1.45$).

All optical constant spectra are shown in Fig.~\ref{fig:nk}.

\subsection{FDTD simulation setup}
\label{sec:fdtd}

Simulations were performed in Lumerical FDTD at normal incidence with the plane
wave propagating along $-z$.
The domain is periodic in $x$ and $y$ (one unit cell of width $\Lambda$);
perfectly matched layers (PML) terminate both $z$ boundaries.
Separate simulations were run for TE ($\bm{E} \parallel \hat{x}$)
and TM ($\bm{E} \parallel \hat{y}$).
A frequency-domain reflection monitor above the source records $\tilde{r}(\omega)$;
absorption is $A = 1 - R$, $R = |\tilde{r}|^2$.
Field monitors ($E_x$, $E_y$, $E_z$) and an absorption power density monitor
record the full electromagnetic cross-section.
For an anisotropic absorbing medium, the volumetric absorption power density is
\begin{equation}
  \Pabs = \tfrac{1}{2}\,\omega\,\varepsilon_0\!\left[
    \mathrm{Im}(\exx)\,|E_x|^2 +
    \mathrm{Im}(\eyy)\,|E_y|^2 +
    \mathrm{Im}(\ezz)\,|E_z|^2
  \right],
  \label{eq:pabs}
\end{equation}
which reduces to the standard $\tfrac{1}{2}\omega\varepsilon_0\,\varepsilon''\,|E|^2$ only
in the isotropic limit.
Conformal mesh refinement is applied at all material interfaces.
Crystal orientations $\alpha$, $\beta$, $\gamma$ label which
$\alpha$-MoO\textsubscript{3} principal axis ([100], [010], [001], respectively)
is aligned with the TE direction; the \ReS{} crystallographic axes co-rotate with
MoO\textsubscript{3} in each configuration.

%% ─────────────────────────────────────────────────────────────────────────────
\section{Results and Discussion}
%% ─────────────────────────────────────────────────────────────────────────────

\subsection{Near-perfect absorption and the role of the ReS\textsubscript{2} grating}

\begin{figure}[htbp]
  \centering
  \includegraphics[width=0.60\textwidth]{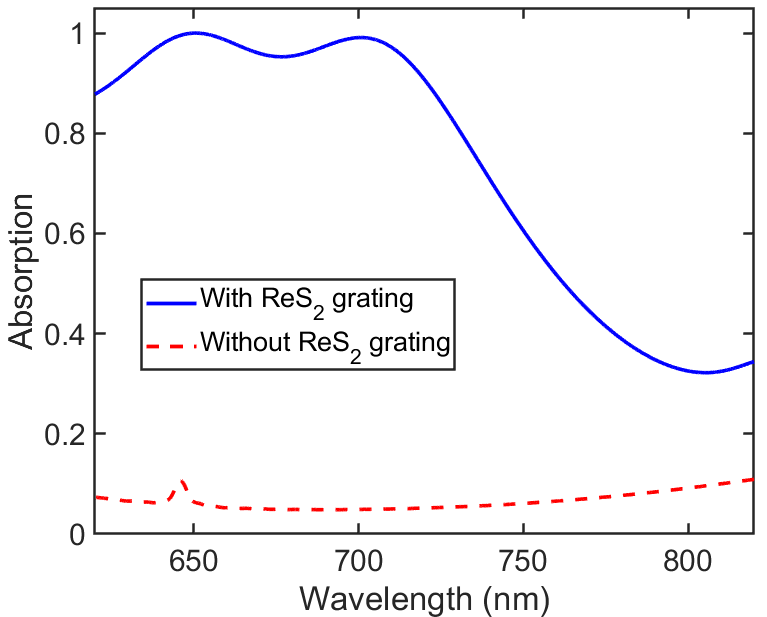}
  \caption{TE absorption spectra with (solid blue) and without (dashed red)
    the \ReS{} stripe grating.
    Period \SI{500}{\nano\meter}, duty cycle 0.5, $\alpha$-orientation.}
  \label{fig:withoutwith}
\end{figure}

Fig.~\ref{fig:withoutwith} compares absorption spectra for the full
heterostructure and an otherwise identical stack from which the \ReS{} grating has
been removed.
With the grating, a sharp TE resonance reaches $A = 99.99\%$ at
$\lres = \SI{650.5}{\nano\meter}$, corresponding to $R = 1.33 \times 10^{-4}$.
Without the grating, the spectrum is featureless and broadband; no resonant feature
survives.
The grating is therefore the resonance-generating element; the \MoO/Au back-reflector
provides the cavity that enables near-zero reflectance.

The near-unity absorption follows from temporal coupled-mode theory
(TCMT)~\cite{Landy2008,Fan2003}.
For a single-mode resonator coupled to one input/output port, with resonance
frequency $\omega_0$, radiative decay rate $\gr$, and intrinsic absorption
rate $\ga$, the absorbed power fraction is
\begin{equation}
  A(\omega) = \frac{4\gr\ga}{\bigl(\omega - \omega_0\bigr)^2 + (\gr + \ga)^2}.
  \label{eq:tcmt}
\end{equation}
At critical coupling ($\gr = \ga$), $A(\omega_0) = 1$.
The resonance quality factor is $Q = \omega_0/(\gr + \ga)$; at critical coupling this
simplifies to $Q = \omega_0/(2\ga)$, set entirely by the absorptive loss of the
\ReS{} mode.
The \ReS{} Mie edge mode supplies a fixed $\ga$ determined by
$\mathrm{Im}(\bm{\varepsilon}_{\mathrm{ReS_2}})$ at the resonance wavelength.
The \MoO{} spacer thickness determines the round-trip phase accumulated in the
cavity formed between the grating and the Au mirror, thereby influencing the
coupling condition. For the optimized geometry, the calculated spectrum exhibits
a narrow resonance near \SI{650.5}{\nano\meter}, as shown in
Fig.~\ref{fig:withoutwith}.

\subsection{Mie-type edge resonance: field and power-density evidence}

\begin{figure}[t]
  \centering
  \begin{subfigure}[b]{0.48\textwidth}
    \includegraphics[width=\textwidth]{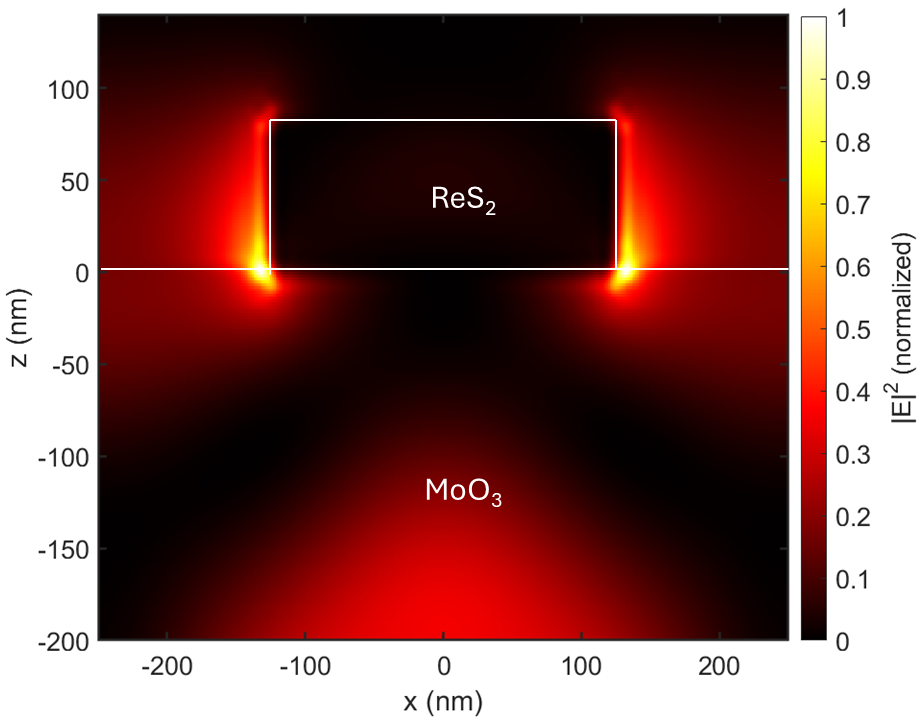}
    \caption{}
    \label{fig:efield_full}
  \end{subfigure}
  \hfill
  \begin{subfigure}[b]{0.48\textwidth}
    \includegraphics[width=\textwidth]{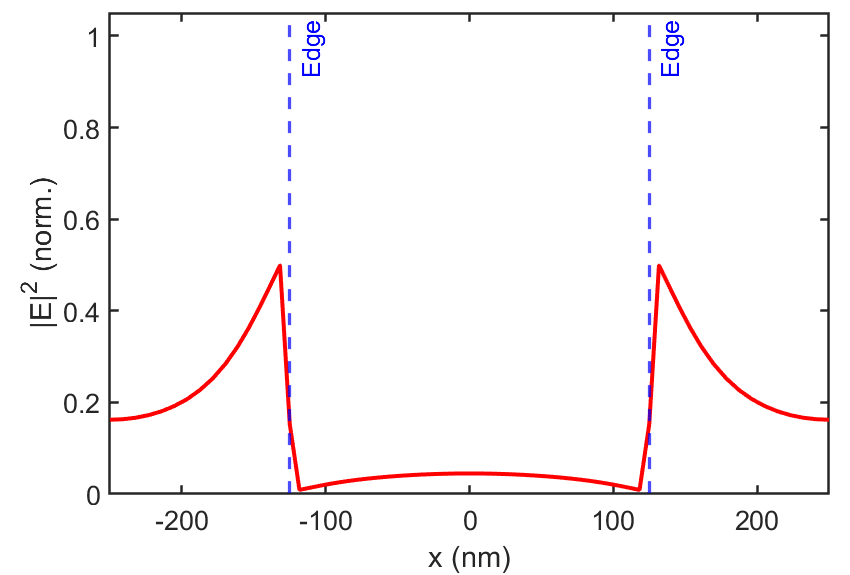}
    \caption{}
    \label{fig:efield_linecut}
  \end{subfigure}
  \caption{$|E|^2$ field distribution at TE resonance ($\lres = \SI{650.5}{\nano\meter}$),
    normalized to incident intensity. (a)~Full $xz$ cross-section; white dashed lines
    indicate the \ReS{} stripe boundaries and the \ReS/\MoO{} interface.
    (b)~Horizontal line cut at mid-stripe height ($z = \SI{40}{\nano\meter}$);
    vertical dashed lines mark stripe edges at $x = \pm\SI{125}{\nano\meter}$.}
  \label{fig:efield}
\end{figure}

\begin{figure}[htbp]
  \centering
  \begin{subfigure}[b]{0.48\textwidth}
    \includegraphics[width=\textwidth]{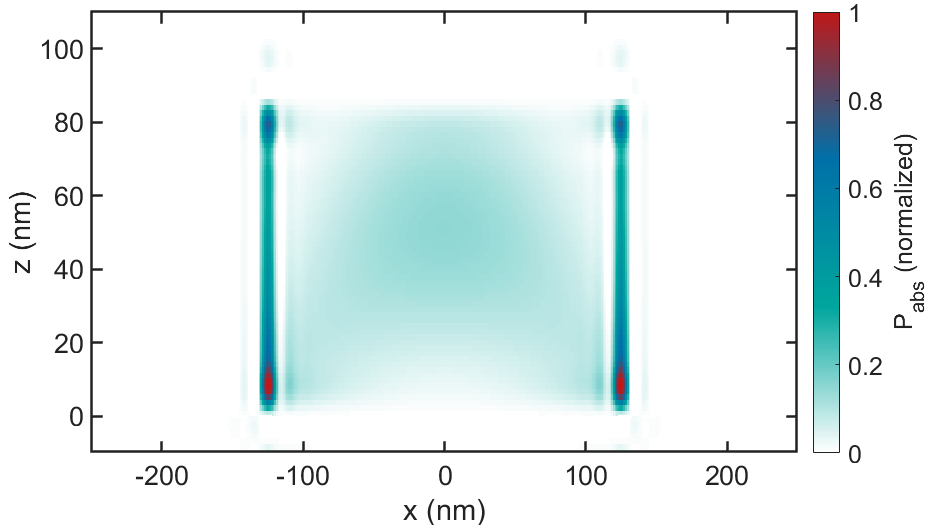}
    \caption{}
  \end{subfigure}
  \hfill
  \begin{subfigure}[b]{0.48\textwidth}
    \includegraphics[width=\textwidth]{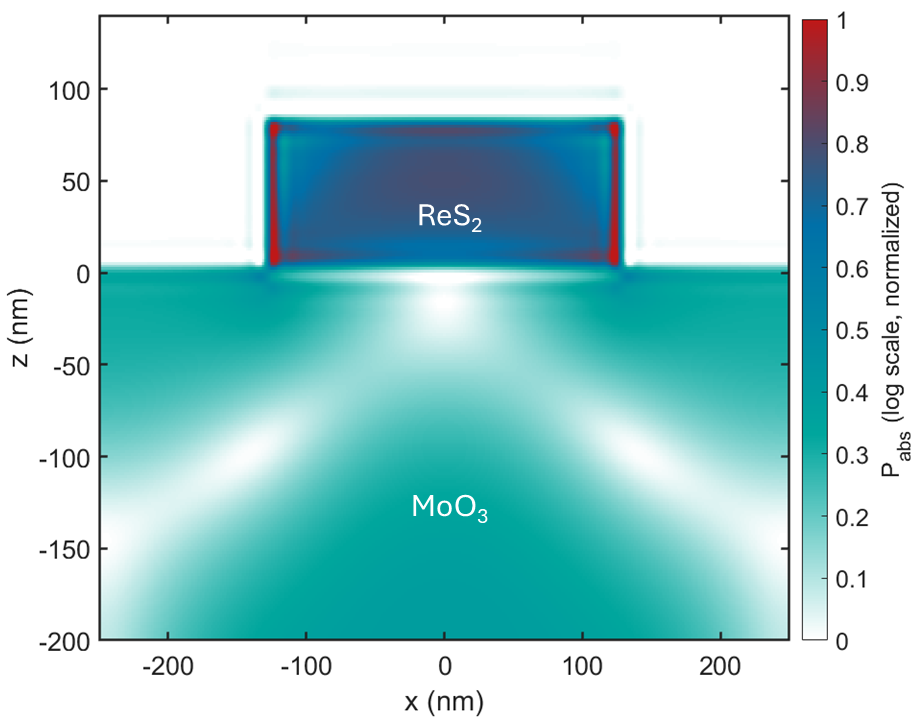}
    \caption{}
  \end{subfigure}
  \caption{Absorption power density $\Pabs$ (Eq.~\ref{eq:pabs}) at TE resonance.
    (a)~Linear scale showing edge-localized dissipation in the \ReS{} layer.
    (b)~Logarithmic scale revealing secondary dissipation into the \MoO{} spacer
    and Au mirror.}
  \label{fig:pabs}
\end{figure}

Figs.~\ref{fig:efield}a--b show the electric-field intensity at the resonance wavelength
($\lambda_{\mathrm{res}}=\SI{650.5}{\nano\meter}$).
The strongest field enhancement occurs near the outer edges of the ReS$_2$ stripe,
whereas the field inside the stripe remains comparatively weak.
The localization is confined to the vicinity of the stripe boundaries rather than
extending across the full grating period, which is consistent with a localized
Mie-type edge resonance.

The absorption power density maps in Fig.~\ref{fig:pabs} exhibit a similar spatial
distribution.
On the linear scale, dissipation is concentrated near the ReS$_2$ stripe edges,
while the logarithmic map reveals weaker absorption within the $\alpha$-MoO$_3$
spacer and the Au back-reflector.
The close correspondence between the $|E|^2$ and $P_{\mathrm{abs}}$ distributions
indicates that absorption occurs primarily in the regions of strongest field
enhancement.

Taken together, the field and power-density maps suggest that the resonance is
dominated by localized electromagnetic confinement near the stripe edges rather
than by a mode distributed throughout the periodic structure.

\subsection{Period dependence and resonance character}

\begin{figure}[t]
  \centering
  \includegraphics[width=0.6\textwidth]{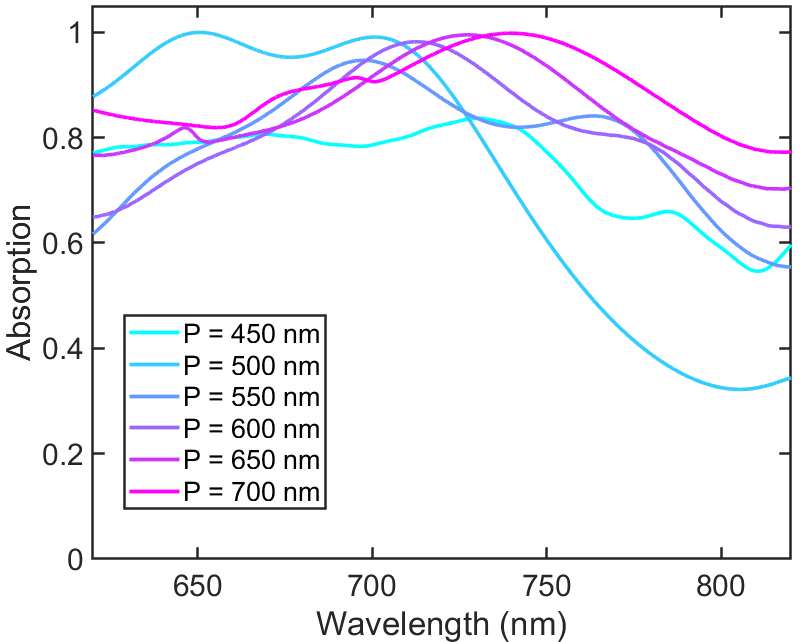}
  \caption{TE absorption spectra for grating periods of 450, 500, 550, 600, 650,
    and \SI{700}{\nano\meter}. Stripe width, \ReS{} thickness, and \MoO{} spacer
    thickness are held constant across all six cases.}
  \label{fig:period}
\end{figure}

\begin{figure}[t]
  \centering
  \includegraphics[width=0.6\textwidth]{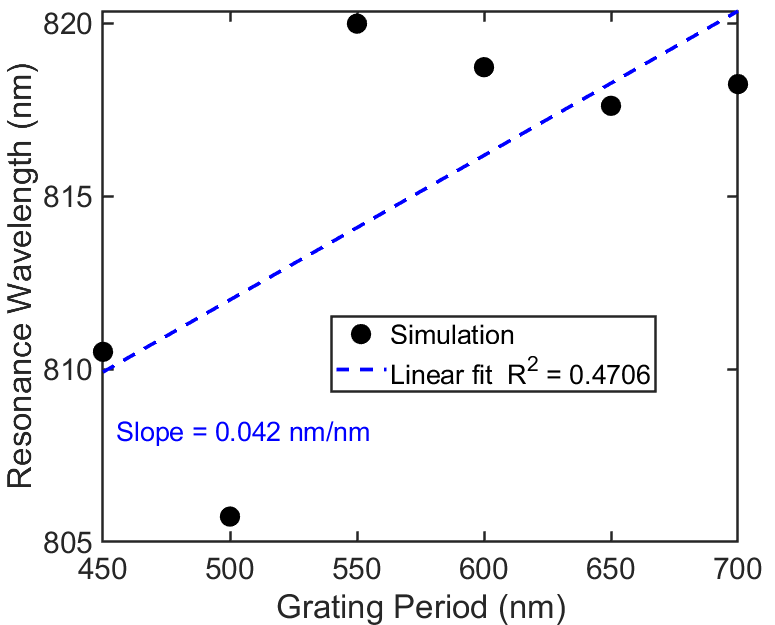}
  \caption{Resonance wavelength $\lres$ versus grating period.
    Closed circles: FDTD data.
    Dashed line: linear fit (slope $= 0.042$~nm/nm, $R^2 = 0.47$).}
  \label{fig:resvperiod}
\end{figure}

Figs.~\ref{fig:period} and \ref{fig:resvperiod} show the effect of grating period
$\Lambda$ on the absorption response while all other geometric parameters are held
constant. As $\Lambda$ increases from 450 to \SI{700}{\nano\meter}, the resonance
wavelength exhibits only a modest shift (Fig.~\ref{fig:period}). A linear fit to
the extracted resonance positions yields a slope of
$d\lambda_{\mathrm{res}}/d\Lambda = 0.042$~nm/nm (Fig.~\ref{fig:resvperiod}).

For a guided-mode resonance, the resonance wavelength is expected to scale more
strongly with grating period through the phase-matching relation
$\lambda_{\mathrm{GMR}} \approx n_{\mathrm{eff}}\Lambda$~\cite{Wang1993}. The weak
dependence observed here suggests that the resonance wavelength is not primarily
determined by the grating periodicity. In addition, the fitted trend captures only
part of the variation in the data ($R^2 = 0.47$), indicating that period alone does
not fully describe the resonance behavior.

The most pronounced effect of increasing $\Lambda$ is visible in the spectral
lineshape (Fig.~\ref{fig:period}), where the resonance depth and linewidth vary
more noticeably than the resonance wavelength. Taken together, these observations
suggest that the resonance is governed mainly by the local geometry and optical
properties of the ReS$_2$ stripe, while the grating period primarily influences the
coupling characteristics of the mode. This behavior is consistent with a localized
Mie-type resonance rather than a guided mode extending across the periodic
structure.

\subsection{Polarization-selective response}

\begin{figure}[t]
  \centering
  \includegraphics[width=0.6\textwidth]{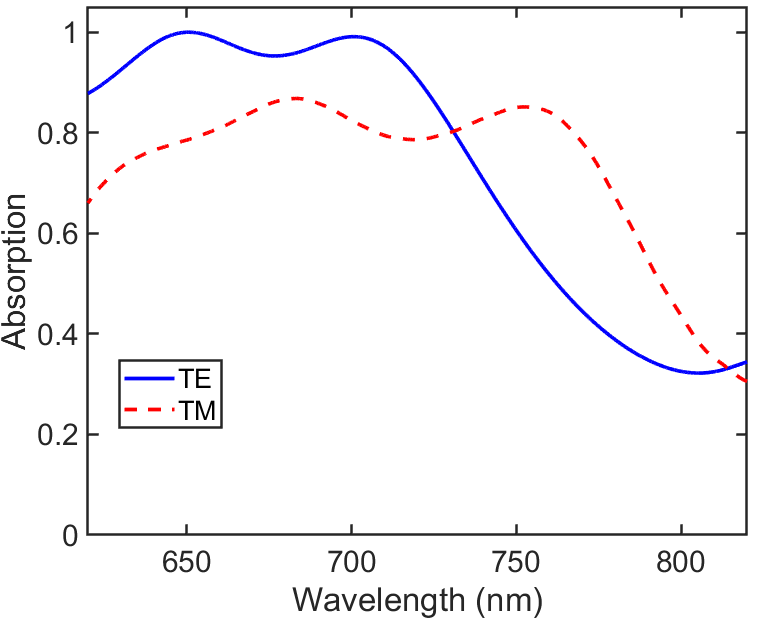}
  \caption{Absorption spectra for TE ($\bm{E}\parallel\hat{x}$, solid blue) and
    TM ($\bm{E}\parallel\hat{y}$, dashed red) illumination.
    Period \SI{500}{\nano\meter}, $\alpha$-orientation, optimized \MoO{} spacer.}
  \label{fig:tetm}
\end{figure}

Fig.~\ref{fig:tetm} shows TE and TM absorption spectra for the optimized geometry.
The resonance splitting is $\Delta\lambda_{\mathrm{TE-TM}} = \SI{16.2}{\nano\meter}$;
the two polarizations also differ in peak absorption and spectral shape.

The origin of the splitting is the biaxial in-plane permittivity of
ReS\textsubscript{2}.
Under TE illumination ($\bm{E}\parallel\hat{x}$), the in-plane electric field in
the stripe body couples primarily to $\exx(\lambda)$, establishing the Mie resonance
condition for the $x$-polarized mode.
Under TM ($\bm{E}\parallel\hat{y}$), it is $\eyy(\lambda)$ that governs the
effective modal index.
Because $\exx \neq \eyy$ at all visible wavelengths (Fig.~\ref{fig:nk}),
with $n_{xx} \approx 4.0 > n_{yy} \approx 3.7$ at \SI{650}{\nano\meter}, the
effective standing-wave condition
\begin{equation}
  n_{\mathrm{eff}}(\lambda)\cdot 2w \approx m\lambda, \quad m \in \mathbb{Z},
  \label{eq:mie_cond}
\end{equation}
is satisfied at different wavelengths for TE and TM, splitting the resonance pair.
This is an intrinsic material mechanism: the stripe geometry is identical for both
polarizations, and no geometric asymmetry is required.

This type of polarization discrimination is available only in materials with
$\exx \neq \eyy$.
The hexagonal $2H$ TMDs---MoS\textsubscript{2}, WS\textsubscript{2},
MoSe\textsubscript{2}, WSe\textsubscript{2}---are in-plane isotropic by their
$D_{3h}$ lattice symmetry, giving $\exx = \eyy$ identically and producing no
intrinsic TE--TM splitting in a stripe of this type~\cite{Ermolaev2021,Munkhbat2022}.
The $1T''$ crystal structure of ReS\textsubscript{2} and the closely related
ReSe\textsubscript{2}~\cite{Shubnic2020,Mikhin2023} breaks this degeneracy,
making the rhenium dichalcogenide family uniquely suited for material-encoded
polarization discrimination in the visible.

\subsection{Isolating the ReS\textsubscript{2} anisotropy contribution}

\begin{figure}[htbp]
  \centering
  \includegraphics[width=\textwidth]{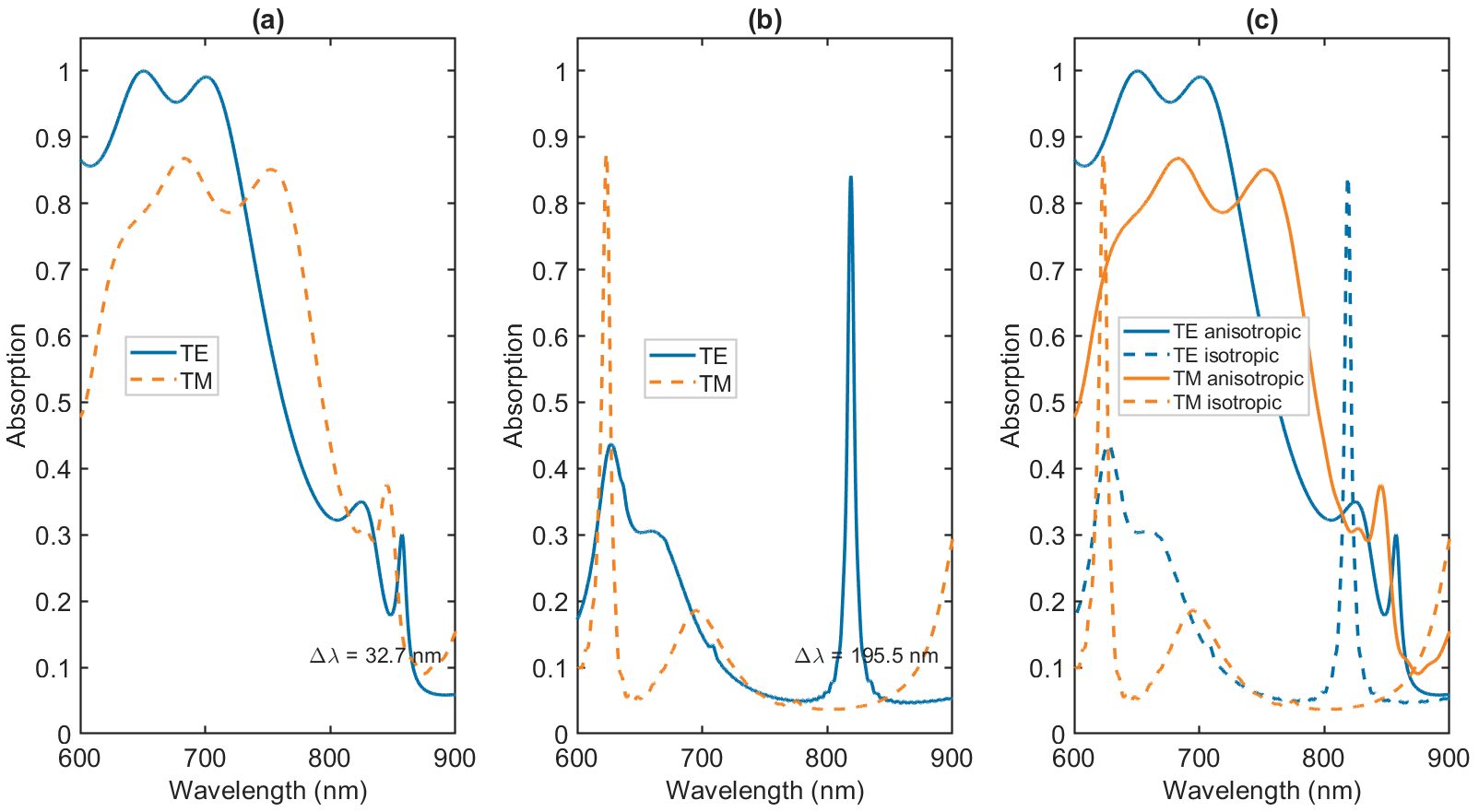}
  \caption{Control simulation replacing biaxial \ReS{} with an isotropic equivalent
    ($n_{\mathrm{iso}} = (n_{xx}+n_{yy})/2$, same extinction).
    Panel (a): anisotropic \ReS{} TE (blue solid) and TM (orange dashed)
    with annotated $\Delta\lambda$.
    Panel (b): isotropic \ReS{} control same polarizations.
    Panel (c): overlay of anisotropic (solid) and isotropic (dashed) for each
    polarization.
    All other parameters unchanged; $\alpha$-MoO\textsubscript{3} spacer remains
    anisotropic throughout.}
  \label{fig:res2iso}
\end{figure}

To isolate the contribution of \ReS{} biaxiality from that of \MoO{} anisotropy,
we replaced the biaxial ReS\textsubscript{2} tensor with an isotropic equivalent
using volume-averaged in-plane indices
$n_{\mathrm{iso}} = (n_{xx}+n_{yy})/2$ and the same extinction coefficients,
while leaving the \MoO{} spacer fully anisotropic.
Fig.~\ref{fig:res2iso} shows the result.

Replacing the biaxial ReS$_2$ tensor with an isotropic equivalent reduces the
separation between the TE and TM resonances (Figs.~\ref{fig:res2iso}a--b).
The spectral responses of the two polarizations also become more similar, although
a residual wavelength difference remains due to the anisotropic $\alpha$-MoO$_3$
spacer. As shown in Fig.~\ref{fig:res2iso}c, isotropizing ReS$_2$ shifts the
resonance positions and reduces the overall polarization contrast. These changes
indicate that the in-plane anisotropy of ReS$_2$ plays an important role in the
observed polarization-dependent response, while the remaining splitting suggests
an additional contribution from the $\alpha$-MoO$_3$ layer.

\subsection{Role of $\alpha$-MoO\textsubscript{3} anisotropy}
\label{sec:moo3}

\begin{figure}[htbp]
  \centering
  \includegraphics[width=\textwidth]{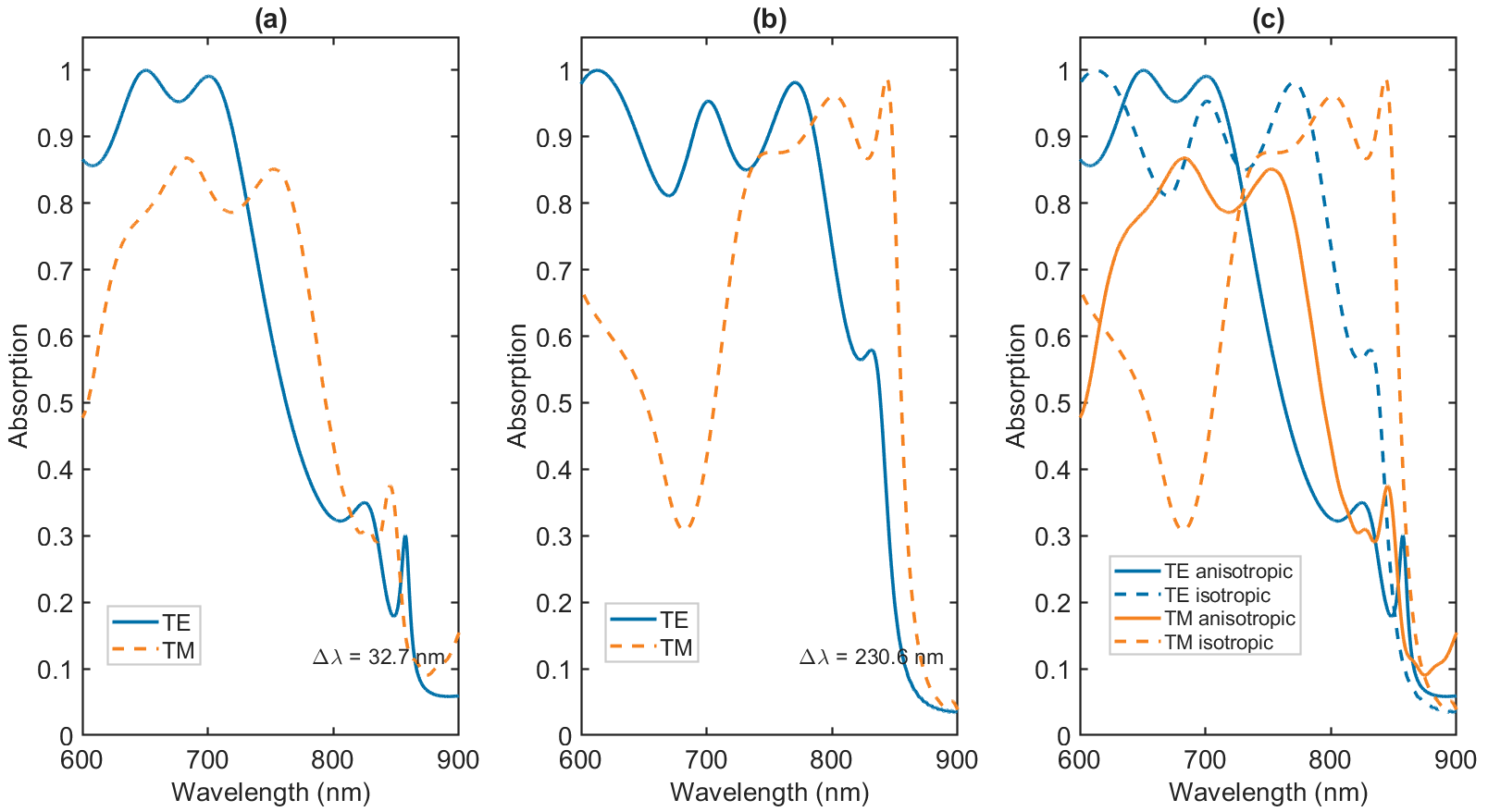}
  \caption{Control simulation replacing anisotropic \MoO{} with an isotropic
    equivalent ($n_{\mathrm{iso}} = 2.44$, $k_{\mathrm{iso}} = 0.012$ at
    \SI{650}{\nano\meter}).
    Panel (a): fully anisotropic \MoO{} --- TE (blue solid) vs TM (orange dashed),
    with annotated $\Delta\lambda = \SI{16.2}{\nano\meter}$.
    Panel (b): isotropic \MoO{} control, showing TM spectral inversion and
    annotated $\Delta\lambda = \SI{2.5}{\nano\meter}$.
    Panel (c): overlay of anisotropic (solid) and isotropic (dashed) cases.
    \ReS{} remains biaxial throughout.}
  \label{fig:moo3iso}
\end{figure}

A second control replaces the anisotropic \MoO{} spacer with an isotropic equivalent
($n_{\mathrm{iso}} = 2.44$, $k_{\mathrm{iso}} = 0.012$ at \SI{650}{\nano\meter}),
while the biaxial \ReS{} layer is kept unchanged. Under TE illumination, the
resonance wavelength shifts slightly and the peak absorption decreases modestly.
The TM response is affected more strongly, with a relative minimum in the
anisotropic case evolving into a peak after isotropization. As a result, the
TE--TM wavelength separation decreases from \SI{16.2}{\nano\meter} to
\SI{2.5}{\nano\meter}.

The mechanism is a polarization-dependent round-trip cavity phase.
In the anisotropic \MoO{} spacer, TE fields propagate along the $\alpha$-axis
($n_\alpha \approx 2.13$) while TM fields see the $\gamma$-axis
($n_\gamma \approx 2.72$) for the $\alpha$-crystal orientation.
The differential round-trip phase accumulated in the spacer is
\begin{equation}
  \Delta\phi
    = \frac{4\pi\,t_{\mathrm{MoO_3}}}{\lambda}
      \bigl(n_\gamma - n_\alpha\bigr)
    \approx \frac{4\pi \times 300}{650.5}\times 0.59
    \approx 3.4\;\mathrm{rad},
  \label{eq:deltaphi}
\end{equation}
where $t_{\mathrm{MoO_3}} = \SI{300}{\nano\meter}$.
This differential phase of $\approx\pi$ means the two polarizations experience nearly
opposing cavity conditions: TE is tuned to critical coupling while TM is driven far
from it at the same wavelength.
In the isotropic substitute, both polarizations accumulate identical round-trip phases
($n_\alpha = n_\gamma = 2.44$), collapsing the differential phase to zero and removing
the cooperative amplification of TE--TM contrast.
The qualitative TM inversion indicates that the TM
critical-coupling condition is not merely shifted in wavelength but fundamentally
disrupted when the cavity phase differential is removed.
The \MoO{} anisotropy therefore functions as an active polarization discriminator
through a phase-engineering mechanism, not as a passive dielectric spacer.

\subsection{Crystal orientation dependence}

\begin{figure}[htbp]
  \centering
  \includegraphics[width=\textwidth]{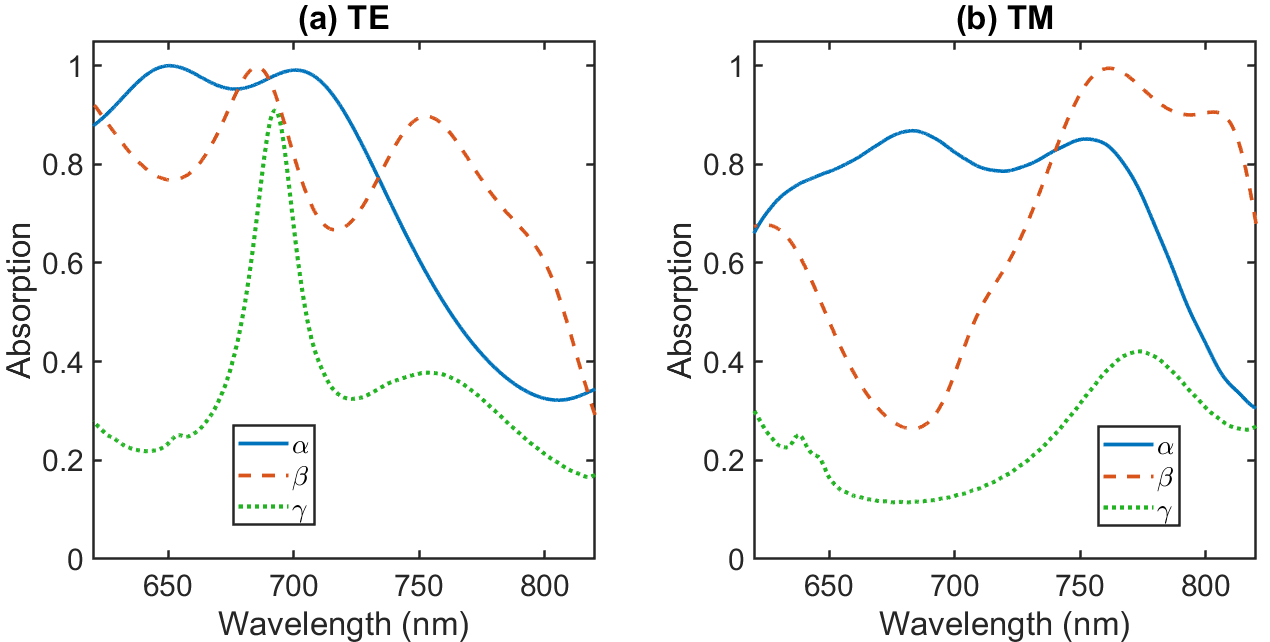}
  \caption{Absorption spectra for three in-plane crystal orientations
    ($\alpha$, $\beta$, $\gamma$) of the dual-anisotropic layer stack.
    Left panel: TE polarization; right panel: TM polarization.
    Grating geometry is identical across all three configurations.}
  \label{fig:orientation}
\end{figure}

Fig.~\ref{fig:orientation} shows absorption spectra when the in-plane crystal
orientation of both the \ReS{} and \MoO{} layers is rotated among the three
principal-axis configurations labeled $\alpha$, $\beta$, $\gamma$.
In the $\alpha$-configuration, the MoO\textsubscript{3} [100] axis is aligned with
the TE polarization direction; in $\beta$, the [010] axis (the stacking, or
interlayer, direction of $\alpha$-MoO\textsubscript{3}) is brought in-plane; in
$\gamma$, the [001] axis is aligned with TE.
The \ReS{} crystallographic axes co-rotate with \MoO\textsubscript{3} in each
configuration.

Both the resonance wavelength and the peak absorption shift systematically across
all three orientations for both TE and TM.
The $\beta$-configuration is particularly noteworthy: bringing the MoO\textsubscript{3}
van der Waals stacking direction in-plane changes the effective cavity permittivity
from $n_\alpha, n_\gamma$ to the $n_\beta$ component (\SI{2.48}{\nano\meter}
at \SI{650}{\nano\meter}), which shifts the cavity phase accumulation and the
critical-coupling match.
The TM response in the $\beta$-configuration achieves near-perfect absorption at a
wavelength near \SI{762}{\nano\meter} (Table~\ref{tab:resonance}), distinct from
both the $\alpha$- and $\gamma$-cases, illustrating how orientation provides
access to a different critical-coupling wavelength without any geometric change.

Since the grating geometry is identical in all three cases, the observed spectral
changes originate from the anisotropic optical properties of the ReS$_2$ and
$\alpha$-MoO$_3$ layers. The orientation dependence shown in
Fig.~\ref{fig:orientation} suggests that both materials contribute to the
polarization-dependent response and provide an additional degree of control over
the resonance characteristics.

\subsection{Performance summary}

Table~\ref{tab:resonance} lists the resonance characteristics for all six
orientation--polarization combinations.
The $\alpha$-orientation, TE polarization achieves the design target of
$A = 99.99\%$ at $\lres = \SI{650.5}{\nano\meter}$.
Near-perfect absorption is also achieved for $\beta$-TE ($99.98\%$) and
$\beta$-TM ($99.97\%$), each at markedly different wavelengths.
The $\gamma$-TM configuration yields the lowest absorption (58.1\%), reflecting a
configuration where the phase-matching between the Mie resonance and the cavity
condition is least favorable.
The range of accessible peak-absorption wavelengths across Table~\ref{tab:resonance}
spans $\approx\SI{120}{\nano\meter}$ (650--772~nm), achievable entirely through
crystal orientation selection and without any re-patterning of the grating.

\begin{table}[htbp]
  \centering
  \caption{Resonance characteristics for all orientation--polarization configurations.
    Orientation labels correspond to which $\alpha$-MoO\textsubscript{3} principal
    axis is aligned with the TE direction; the \ReS{} axes co-rotate.}
  \label{tab:resonance}
  \smallskip
  \begin{tabular}{llccc}
    \toprule
    Orientation & Polarization
      & $\lres$ (\si{\nano\meter})
      & $R_{\min}$ (\%)
      & $A_{\max}$ (\%) \\
    \midrule
    $\alpha$ & TE & 650.5 & 0.01 & 99.99 \\
    $\alpha$ & TM & 688.3 & 12.6  & 87.4  \\
    $\beta$  & TE & 689.7 & 0.02  & 99.98 \\
    $\beta$  & TM & 761.6 & 0.03  & 99.97 \\
    $\gamma$ & TE & 692.3 & 0.9   & 99.1  \\
    $\gamma$ & TM & 772.1 & 41.9  & 58.1  \\
    \bottomrule
  \end{tabular}
\end{table}

\begin{table}[htbp]
  \centering
  \caption{Comparison with related resonant absorber and TMD photonic devices.}
  \label{tab:comparison}
  \smallskip
  \footnotesize
  \setlength{\tabcolsep}{4pt}
  \resizebox{\textwidth}{!}{%
  \begin{tabular}{@{}llllccc@{}}
    \toprule
    Reference
      & Material system
      & Mechanism
      & $A_{\max}$
      & Polarization selectivity
      & $\lambda$ range \\
    \midrule
    Verre et al.\ \cite{Verre2019}
      & WS\textsubscript{2} nanodisk
      & Mie dipole (scattering)
      & ---
      & Geometric
      & 550--700~nm \\
    Green et al.\ \cite{Green2020}
      & TMD nanodisk
      & Mie + material anisotropy
      & ---
      & Uniaxial out-of-plane
      & 550--800~nm \\
    Munkhbat et al.\ \cite{Munkhbat2019}
      & TMD multilayer
      & Self-hybridized polariton
      & $\sim$85\%
      & Limited
      & 600--800~nm \\
    Tang et al.\ \cite{Tang2021}
      & $\alpha$-MoO\textsubscript{3}/Ag nanoribbon
      & Plasmonic--dielectric
      & $\approx$100\% (3-band)
      & Geometric
      & 400--700~nm \\
    Yan et al.\ \cite{Yan2023}
      & \ReS/WSe\textsubscript{2}
      & Mie resonance
      & Detection only
      & Biaxial (material)
      & 600--800~nm \\
    \textbf{This work}
      & \textbf{\ReS/\MoO/Au}
      & \textbf{Mie edge $+$ critical coupling}
      & $\bm{99.99\%}$
      & \textbf{Dual biaxial material}
      & \textbf{650--772~nm} \\
    \bottomrule
  \end{tabular}%
  }
\end{table}

Table~\ref{tab:comparison} situates the present device in the recent literature.
Previous TMD resonator work has achieved polarization contrast through either
geometric meta-atom asymmetry or the out-of-plane birefringence common to all
van der Waals materials~\cite{Green2020,Ermolaev2021}.
The combination of in-plane biaxiality at the grating resonator level (from \ReS)
with anisotropy at the cavity spacer level (from \MoO), yielding both near-perfect
absorption and materials-controlled orientation tunability, does not appear in prior
reports.

\subsection{Fabrication considerations}

The proposed heterostructure is assembled from process steps that have each been
demonstrated individually in the van der Waals photonics literature.
The \MoO/Au back-reflector is deposited by sequential thermal evaporation of
Au (\SI{150}{\nano\meter}) and $\alpha$-MoO\textsubscript{3} (\SI{300}{\nano\meter})
onto a SiO\textsubscript{2} substrate; both materials are commercially available
and routinely deposited by e-beam or thermal evaporation.
The \ReS{} grating layer begins with mechanical exfoliation of ReS\textsubscript{2}
from a bulk crystal onto a polydimethylsiloxane (PDMS) stamp, followed by
deterministic dry-transfer onto the \MoO{} surface~\cite{Castellanos2014}.
The critical step prior to transfer is identification and alignment of the
ReS\textsubscript{2} crystallographic axes: in-plane biaxial anisotropy makes
the $b$-axis orientation directly accessible via polarimetric Raman
spectroscopy~\cite{Aslan2016} or scattering-type scanning near-field optical
microscopy~\cite{Mooshammer2022}, both of which resolve the principal axes to
within a few degrees.
The \ReS{} stripe grating is then defined by focused-ion-beam (FIB) milling,
which has been used to pattern WS\textsubscript{2} and ReS\textsubscript{2}
nanostructures with sub-20~nm edge
precision~\cite{Verre2019,Yan2023,Munkhbat2020}.
Alternatively, electron-beam lithography followed by reactive-ion etching
(EBL/RIE) in SF\textsubscript{6}/O\textsubscript{2} chemistry provides
wafer-scale patterning with comparable edge quality.

The weak dependence of resonance wavelength on grating period suggests a greater
tolerance to period variations than is typically expected for guided-mode
resonance structures. As shown in Fig.~\ref{fig:resvperiod}, changes in
$\Lambda$ produce only modest shifts in the resonance position, while the
dominant effects are observed in the resonance depth and linewidth. This behavior
may be beneficial for large-area fabrication approaches, where small variations
in grating period are difficult to avoid.

\section{Conclusion}

We have investigated a polarization-selective absorber consisting of a biaxial
\ReS{} stripe grating, an anisotropic \MoO{} spacer, and an Au back-reflector
using finite-difference time-domain simulations. The optimized structure
achieves near-perfect absorption ($A = 99.99\%$,
$R = 1.33 \times 10^{-4}$) at
$\lambda = \SI{650.5}{\nano\meter}$ under TE
illumination. In addition, crystal-orientation control enables resonance tuning
over a wavelength range of approximately \SI{120}{\nano\meter} without modifying
the device geometry.

Field distributions, absorption power density maps, and period-dependent studies
indicate that the observed resonance is localized near the edges of the
\ReS{} stripe and exhibits only a weak dependence on grating period.
Control simulations further show that the polarization-dependent response
originates primarily from the in-plane anisotropy of ReS\textsubscript{2},
while the anisotropic \MoO{} spacer strongly influences the spectral separation
between TE and TM resonances. Together, these effects produce the pronounced
polarization-selective absorption observed in the optimized structure.

Although the present study focuses on the \ReS/\MoO{} material system, the
underlying concept may be extended to other combinations of anisotropic
resonator and spacer materials. The results highlight the potential of material
anisotropy as a design parameter for controlling resonant absorption and
polarization response in van der Waals photonic structures. Future work may
explore fabrication tolerances, angular performance, and active tuning through
electrostatic gating or applied strain.

\begin{backmatter}

\bmsection{Acknowledgments}
The authors thank the Department of Electrical and Electronic Engineering, Bangladesh
University of Engineering and Technology (BUET) for computational resources.

\bmsection{Disclosures}
The authors declare no conflicts of interest.

\bmsection{Data Availability Statement}
All key structural parameters and simulation conditions required to reproduce the reported results are provided within the manuscript.
Additional simulation data and supporting materials are available from the corresponding authors upon reasonable request.

\end{backmatter}

\bibliography{references}

\end{document}